# Highly Efficient Fuel Cell Electrodes from Few-Layer Graphene Sheets and Electrochemically Deposited Palladium Nanoparticles

*Michael Höltig, Charlotte Ruhmlieb, Tobias Kipp\* and Alf Mews*

Institute of Physical Chemistry, University of Hamburg, Grindelallee 117, 20146 Hamburg, Germany

ABSTRACT

An extremely efficient ethanol fuel cell electrode is produced by combining the large surface area of vertically oriented and highly conductive few-layer graphene sheets with electrochemically deposited palladium nanoparticles. The electrodes show an extraordinary high catalyst activity of up to 7977 mA/(mg Pd) at low catalyst loadings of 0.64 μg/cm² and a very high current density of up to 106 mA/cm² at high catalyst loadings of 83 μg/cm². Moreover, the low onset potentials combined with a good poisoning resistance and long-term stability make these electrodes highly suitable for real applications. These features are achieved by using a newly developed electrochemical catalyst deposition process exploiting high voltages of up to 3.5 kV. This technique allows controlling the catalyst amount ranging from a homogeneous widespread distribution of small (≤ 10 nm) palladium nanoparticles to rather dense layers of particles, while every catalyst particle has electrical contact to the graphene electrode.

**INTRODUCTION**

Direct alcohol fuel cells belong to the most promising candidates for stationary and mobile low-temperature energy production.[1,2] In comparison to gaseous fuels like hydrogen, liquid alcoholic fuels like methanol and ethanol have the advantage of an easier and safer usage and transport. Alcohol fuel cells convert the chemical energy stored in small alcohol molecules into electricity. However, so far methanol cannot be considered as a candidate for high performance energy generation in fuel cells because of its toxicity and the so-called methanol crossover.[3,4] Furthermore methanol oxidation leads to the formation of carbon monoxide which can possibly poison the catalyst. Ethanol on the other hand is not as toxic and has a volume energy capacity of 6.3 kWh/L, which is even higher than the values for methanol (4.8 kWh/L) and hydrogen (2.6 kWh/L).[5] In general, a metal-catalyzed ethanol-oxidation reaction (EOR) in direct ethanol fuel cells (DEFCs) can take place in acidic or in alkaline media.[6] The EOR in acidic solution needs a considerable amount of scarce and expensive platinum, which limits the commercialization of acid-type DEFCs.[7] The alkaline oxidation of ethanol is an alternative and more economic pathway since it allows the usage of other, less noble metal catalysts than platinum.[8] It has been found that Pd catalysts exhibit higher activities and poisoning resistances than Pt in alkaline solutions.[2,9–11] Several research groups are working on electrochemical systems promoting the EOR in alkaline solution in order to maximize current density and catalyst activity. For example, in the year 2009 Liang *et al.* reported on the EOR on a palladium-disk electrode in 1 M potassium hydroxide (KOH) and 1 M ethanol solution.[12] Their results display a maximum current density of almost 56 mA/cm$^2$. Three years later Liu *et al.*[13] achieved a maximum current density of 90 mA/cm$^2$ in 1 M KOH + 1 M ethanol by using reduced graphene oxide loaded with Pd-Ag-bimetallic nanoparticles. The promising combination of conductive carbon and palladium catalyst particles was further developed and utilized. In 2013 Sawangphruk *et al.* designed a system consisting of an ultra-porous arrangement of Pd nanocrystals, electrochemically deposited on reduced graphene oxide and carbon fiber paper (Pd/rGO/CFP). Using a 0.5 M sodium hydroxide + 1 M ethanol solution they reached a catalyst activity of up to 1033 mA/(mg Pd),[14] which corresponds to a current density of 413.2 mA/cm². Albeit the shown systems and their reached values are not plenary comparable because of different experimental parameters, they show some contributing factors for a powerful EOR system. It turns out that a combination of highly conductive carbon nanostructures with a large surface area as a support for nanoscopic catalysts is very promising for the EOR. Ideally, the nanoparticle catalyst should be homogeneously distributed on the carbon support in order to maximize the catalytic surface area and thereby the catalyst activity. Moreover, the nanoparticle catalyst should have a direct electric contact to its conductive substrate in order to minimize the electric resistance of the system and thereby increase the current density. Especially if the important parameters such as structure and resistance of the support as well as the catalyst morphology and loading can be systematically optimized, advanced fuel cell systems can be achieved. Hence in this work we used an alternative carbon support, namely few-layer graphene (FLG) sheets, which are grown on gold-coated heat-resistant aluminosilicate glass by plasma-enhanced chemical vapor deposition (PECVD). Recently we have shown that these highly conductive carbon nanostructures can be prepared with a very high specific surface area.[15] Here, we used a high-voltage electrochemical deposition technique to prepare Pd nanoparticles directly on the FLG sheets to ensure that every catalyst particle has electrical contact. This method led to controllable catalyst loadings showing a homogenous arrangement of small particles, which is unprecedented in literature so far. Upon systematic optimization of several reaction parameters we have achieved a notably high maximum current density of 106.11 mA/cm² and a record-high maximum catalyst activity of 7977 mA/(mg Pd) for the EOR in 1 M KOH + 1 M ethanol.

**EXPERIMENTAL**

**Synthesis of FLG sheets:** Aluminosilicate glass (20 x 20 mm, 0.7 mm thick) coated with a 50 nm thick gold layer for high electrical conductivity served as substrate material. The synthesis was performed in a custom-built capacitively coupled radio-frequency (13.56 MHz) plasma-enhanced CVD (CC-PECVD) setup.[15] A gas mixture of $C_2H_2$, $H_2$, and He with flow rates of 10, 80, and 70 standard cubic centimeters per minute (sccm), respectively, were used. The substrate temperature of 650 °C was controlled by a resistively heatable stage that also acts as counter electrode. The plasma power and the total pressure were set to 30 W and 11-12 mbar, respectively. The reaction time was 60 min. More details on the synthesis of the carbon structures and their characterization via scanning electron microscopy and Raman spectroscopy, which proves the structures to be FLG sheets, can be found in Ref. 15.

**Electrochemical deposition of palladium nanoparticles:** Before electrochemical deposition, the carbon nanostructures have been electrically contacted by conducting silver paint and copper stripes. These contacts were then isolated with a polymer glue to avoid contamination of the electro chemical reaction solutions. The contacted substrates were mounted into an electrochemical cell that allowed for an exact adjustment of the distance between the sample and the palladium counter electrode between 5 and 11 mm. Pure methanol or a mixture of methanol and chloroform was used as an electrolyte in which the palladium deposition was accomplished by applying short (~ 6 μs) high-voltage pulses of 3.5 kV with frequencies of about 20 Hz for 7 - 8 min. The technical details of the electrochemical set-up will be described elsewhere. After Pd deposition the samples were rinsed with methanol and water and were dried by nitrogen flow.

**Preparation of the reference samples:** Two kinds of reference samples have been prepared. For the reference sample of the first kind, instead of an electrochemical deposition, a layer of Pd with a nominal thickness of 6 nm has been sputtered onto the FLG sheets with a sputtering device (PECS Model 682, Gatan). For the reference samples of the second kind, instead of the FLG sheets, a layer of 5 or 10 nm carbon has been sputtered (PECS Model 682, Gatan) onto the gold coated substrate before Pd nanocrystals had been electrochemically deposited.

**Scanning electron microscopy (SEM) characterization:** SEM images were obtained using a FEI Quanta 3D FEG with an acceleration voltage of 5-20 kV.

**Cyclic voltammetry:** For cyclic voltammetry a potentiostat from Iviumstat controlled by the software Iviumsoft was used. For all measurements a scan rate of 100 mV/s within a potential range between -0.8 V and 0.1 V vs. SCE was set. Before starting measurements the solution was purged with nitrogen for 10 min. The concentrations of solutions were 1 M for KOH and ethanol. We have tested the stability and thus suitability of the SCE in alkaline media by monitoring anodic peak potentials in cyclic-voltammetry experiments over time and compared them to values obtained with an Ag/AgCl reference electrode. Both electrodes behave similar and show no sign of degradation in 1 M KOH and ethanol solution. All measurements were performed at room temperature. If not otherwise stated, for each measurement at least 50 cycles were recorded and the voltammograms shown in this work represent waves between cycle number 45 and 50. For the determination of current densities the substrate area on which the FLG sheets were grown was measured to be 2.8 $cm^2$.

**Determination of catalyst loading:** The determination of the palladium loading is based on the absolute amount of deposited Pd. After electrochemical characterization, the copper stripes, silver paint, and insulation glue, which are necessary for contacting purposes, were mechanically detached from the samples. The samples were then solubilized in aqua regia to dissolve the palladium. The palladium solutions were then filtered to remove potential carbon flakes and subsequently analyzed via ICP-OES (SpectroCiros, Spectro).

**RESULTS AND DISCUSSION**

Figure 1 and 5a show scanning electron microscopy (SEM) images of PECVD-grown FLG-sheet structures covered by electrochemically deposited Pd nanoparticles with different loadings in the range of 0.64 - 83.32 µg Pd/cm² related to the geometrical area of the electrode. From Figure 1a and 1b one can deduce that for low Pd loadings, nanoparticles are nearly homogeneously distributed on the FLG sheets not only at the upper edges but also on the vertical faces. For a Pd loading of 13.71 µg/cm$^2$ (see Figure 5a), the sheets are completely covered by a dense layer of small nanoparticles. The Pd particle size estimated by SEM is 10 nm or less and it seems to be independent of the Pd loading. Increasing the Pd loading leads to a surface lamination by stacking of particle layers (see Figure 1c) before shadowing effects lead to the formation of larger and rounded Pd structures on the top of the FLG support (see Figure 1d). The SEM images in Figure 1 and 5a have been obtained before the samples' catalytic properties with respect to the EOR in alkaline solution have been investigated by cyclic voltammetry, chronoamperometry and further comparative SEM.

Figure 2 shows cyclic voltammograms of the electro-oxidation of ethanol in alkaline solution measured for six FLG-sheet samples with different catalyst loadings between 0.64 - 83.32 µg Pd/cm². Depicted is the current density vs. the potential with respect to the saturated calomel electrode (SCE). The voltammograms exhibit two peaks. The right one, between -0.3 and +0.1 V, represents the forward anodic peak. Its maximum position shifts in positive potential direction and its height is increasing from 5 mA/cm$^2$ to 106 mA/cm$^2$ for increasing Pd loadings. A backward anodic peak occurs between -0.2 and -0.4 V. This peak is also increasing and shifting in positive direction with increasing Pd loading.

The exact reaction mechanism for the electro-oxidation of ethanol in alkaline solution is still under discussion.[2,5,11,14,16,17] It is assumed that during the reaction $CH_3CO$ is adsorbed on the catalyst as an intermediate while the oxidation stops after the formation of acetic acid ($CH_3COOH$). $CO_2$ is not generated within the used potential range. Within this concept, the forward anodic peak can be assigned to the adsorption and partial oxidation of the ethanol on the surface of the Pd. The emerging intermediate $CH_3CO_{ads}$ can poison the catalyst. The backward anodic peak in turn can then be attributed to the stripping of the poisoning intermediate.[14] Another explanation attempt for the occurrence of both peaks assumes the formation of PdO on the Pd surface at high potentials (forward anodic peak) which leads to a poisoning due to the strong binding of intermediates to PdO and therefore a drop of catalyst activity. The fast onset of the backward anodic peak then results from the reduction of the thin PdO layer that should lead to a rapid desorption of the intermediates and the formation of a fresh, catalytically active Pd surface.[5]

The forward anodic peak potentials as well as the corresponding maximum current densities, which are key measures of the voltammograms shown in Figure 2, are compiled in Table 1 where they can be compared to other values as derived in the following. The three bottom rows in Table 1 represent values obtained for the reference samples, which either contain (i) sputtered Pd instead of electrochemically deposited Pd or sputtered carbon, (ii) 5nm and (iii) 10 nm in thickness, instead of FLG sheets. It can be seen that the maximum current density of the sputtered-Pd sample is considerably smaller than for samples with electrochemically deposited Pd of similar loadings on FLG sheets. This is due to a larger contact resistance of sputtered material compared to electrochemically deposited material, for which a direct electric contact to the electrode is inherent. The sputtered-carbon samples also reveal considerably smaller maximum-current densities as compared to FLG-samples with similar Pd loading. This is because of a higher resistance of a sputtered carbon layer compared to the FLG sheets.

For a further analysis of the measurements, the voltammograms can be depicted in terms of the catalyst activity. Therefore the current-density traces, as given in Figure 2, are divided by the corresponding catalyst loading. Results are shown in Figure 3. It can be seen that the catalyst activity is decreasing with increasing Pd loading. The values of the activity determined from the forward anodic peaks are also given in Table 1. They are in each case higher than the value for the sputtered-Pd reference sample and, for comparable Pd loadings, also for the sputtered-carbon reference samples. For the lowest Pd loading of 0.64 µg/cm² a very high activity of 7977 mA/(mg Pd) is achieved, which is to our knowledge the highest published value for pure Pd in the electro-oxidation of ethanol in alkaline solution. We assume that this activity is reached due to the homogeneous distribution of the catalyst that ensures a very good chemical accessibility of each Pd nanoparticle. Additionally, regarding the carbon-Pd interface, it is beneficial that the active catalyst particles are strongly bound to the crystalline few-layer graphene sheets due to donor-acceptor interactions as described in Ref. 18. Electron density from the support is transferred to the 5s orbitals of Pd while electron density of the 4d Pd orbitals is donated to the contiguous carbon, leading to extraordinary electron-transfer properties and a stabilization of the catalytic active ionic state of Pd.[18] By increasing the Pd loading to 83.32 µg/cm² the catalyst activity decreases to 1274 mA/(mg Pd). This decrease is a result of the increase in particle density leading to the formation of dense multilayers and therefore to a smaller active surface area per particle. The overall catalytic surface area however is increasing with increasing Pd loading, explaining the increase in the current density.

The increasing forward anodic peak potential together with the decreasing catalyst activity for increasing catalyst loadings hint to a strong activation effect of the carbon on the catalyst by delivering electrons to the Pd. It is well-known that especially graphene can increase the reactivity of metal particles on its surface via orbital interaction.[19,20] This effect occurs on a short spatial range. It leads to a decreased oxidation potential necessary for the catalytic reaction. With increasing Pd loading, however, the distance between the carbon atoms and the active Pd surface increases and weakens the interaction effect and thus the forward anodic peak potential increases.

The voltammograms in Figure 2 can be further analyzed by integrating the current density of both the forward as well as the backward anodic peak over the time and by subsequently calculating the ratio $I_f / I_b$ of both integrals. Results are compiled in Table 1. The $I_f / I_b$-ratio of our samples increases with increasing Pd loading, starting from 1.36 to 4.04 and 3.69 for the highest loadings, where the latter value is limited by the limited voltage-scan range that cannot record the full forward anodic peak. The reference samples exhibit ratios between 1.00 and 1.55, values considerably lower than for the FLG-sheet-based samples with electrochemically deposited Pd of comparably high loadings. Since the backward anodic peak is linked to the poisoning of the catalyst by adsorbed intermediates, a lower intensity of this peak hints to a more complete reaction during the forward oxidation. Therefore a high $I_f/I_b$ –ratio, like observed in our measurements, is an indicator for a high poisoning tolerance.[14,19,21]

In order to investigate the long-term stability of the catalyst, a combination of chronoamperometry, with a constant potential applied, and cyclic voltammetry was used. These experiments were performed for the sample with a loading of 13.71 µg Pd/cm² (cf. Figure 5a). First, the catalyst of the previously air-dried sample has been refreshed by 10 cycles of voltammetry under the same conditions as in the previously performed cyclic-voltammetry experiments described above. Then a constant voltage of -0.3 V vs. SCE was applied and the catalyst activity was measured for 50 min. The red solid curve in Figure 4a represents the corresponding chronoamperometry trace. It can be well fitted by a biexponential decay with decay times of about 44 s and 930 s, respectively (red dotted curve). The fit asymptotically approaches a catalyst activity of 236 mA/(mg Pd) (red dashed baseline).

After the first chronoamperometry run the catalyst has been refreshed again by 10 cycles of cyclic voltammetry. The trace of the 10[th] cycle is shown in Figure 4b as a red curve. It reaches an activity of more than 3100 mA/(mg Pd), which is still about 85% of the maximum activity measured in the actual cyclic voltammetry measurements discussed before (cf. Table 1). A subsequently recorded chronoamperometry trace (blue solid curve in Figure 4a) is almost identical to the first trace. A biexponential fit (blue dotted curve) yields slightly smaller decay times of 11 s and 857 s while it approaches a slightly lowered baseline at 226 mA/(mg Pd) (blue dashed line). Also the subsequently measured 10[th] cycle of the voltammogram (Figure 4b, blue curve) nearly coincides with the other trace. The non-monoexponential chronoamperometry traces hint to a complex poisoning mechanism that cannot be completely clarified at the moment. However, more importantly, the long decay times in the range of 15 min, the high value of the baseline catalyst activity of about 230 mA/(mg Pd), and the reproducibility of both the chronoamperometry and the refreshing cyclic voltammetry traces prove an excellent poisoning tolerance of the Pd-FLG sheet structure as well as its high long-term stability.

The same sample that has been characterized by chronoamperometry has finally again been imaged by SEM to monitor possible changes of the catalyst appearance. Figure 5b compares an SEM image of the sample after all electrochemical measurements to the already discussed image in Figure 5a that was obtained directly after deposition of the palladium. Obviously, gaps have appeared in the Pd-particle layer, but the overall integrity of the sample is preserved. We want to stress that during all cyclic-voltammetry measurements on all samples with electrochemically deposited Pd on FLG sheets, an increase in the maximum current density with each cycle was observed (not shown here). This is not the case for the reference samples with sputtered carbon instead of FLG sheets; here, the maximum current densities decrease already after the $20^{th}$ ($40^{th}$) cycle for the sample with 10 nm (5 nm) carbon. We deduce that the flat FLG surface is beneficial for a tight attachment of Pd particles such that no appreciable amounts of palladium are detached from the FLG surface. This suggests that the small structural changes, which can be observed in the SEM images of Figure 5, can be attributed to particle rearrangement processes during the electrochemical characterization measurements, which do not derogate the catalytic properties of the Pd-FLG sheet structures.

## CONCLUSIONS

We investigated the catalytic properties of Pd nanoparticles on FLG sheets with respect to the direct ethanol-oxidation reaction in alkaline solution. The FLG sheets have been grown by PECVD on gold-coated aluminosilicate glass substrates. The textured carbon structures provide a very high specific surface area. A high-voltage electrochemical-deposition method was employed to homogenously cover the FLG sheets with Pd nanoparticles. By controlling the amount of deposited Pd, the degree of nanoparticle coverage could be tuned from widespread individual particles over a closely-packed monolayer of particles to a lamination of the carbon support with several layers of nanoparticles.

For such samples with different Pd loadings, the current density for the EOR in alkaline solution has been analyzed via cyclic-voltammetry measurements in 1M KOH + 1M ethanol. A considerably high current density of 106 mA/cm² of the forward anodic peak occurs for the sample with the highest Pd loading. In terms of the catalyst activity the sample with the lowest Pd loading yields an exceptionally high value of 7977 mA/(mg Pd) with a low forward anodic peak potential of -0.295 V vs. SCE.

The current densities for the Pd-FLG sheet structures fabricated by electrochemical deposition are generally much larger than for the reference FLG sample with sputtered Pd of similar loading. This is because of the good electrical contact between the conducting carbon support and the Pd, which is inherent for the electrochemical deposition technique. The current densities are also larger than for the corresponding reference samples with sputtered carbon instead of FLG sheets, which is because of the larger electrical resistance of the sputtered carbon. In a same way, also the catalyst activities of the electrochemically fabricated Pd-FLG sheet samples are much larger than for the sputtered-Pd reference sample of similar loading. This is due to the homogeneous distribution of the Pd nanoparticles that ensures a very good chemical accessibility of each catalyst particle. This is also the reason, why the sputtered-carbon reference sample, despite showing rather small current densities, still exhibit quite high catalyst activities.

Comparatively large values of the $I_f/I_b$ –ratio, i.e. the ratio of the time integrals of the current density of the forward and backward anodic peak, indicate a high poisoning tolerance of the Pd-FLG sheet catalyst structures. Furthermore, also chronoamperometric measurements repeated after a refreshing of the catalyst by cyclic voltammetry prove an excellent poisoning tolerance as well as an exceptional long-term stability of the Pd-FLG sheet structures.

For a technical application of the Pd-FLG catalyst structure in the direct EOR, high current densities as well as high catalyst activities are desirable. However both features are mutually dependent and a large value for one feature typically comes along with a lower value for the other. This behavior is summarized in Figure 6, in which the red and blue data points give the current density and catalyst activity of the samples with different Pd loading, respectively. Dashed lines connect the data point as guides to the eyes; dotted lines are more realistic fits through the data points. A good measure for the most useful Pd loading in view of a maximum value for both current density and catalyst activity is the product (CD × CA) of both values. This product is depicted in Figure 6 as black squares. The corresponding dashed line again connects the data points as a guide to the eye. Additionally, the black dotted line represents the product of the fits of current density and catalytic activity. The depiction of CD × CA suggests an optimal catalyst loading in the range of 20-40 µg Pd/cm² for our system, which corresponds to a catalyst loading of more than one monolayer of palladium nanoparticles.

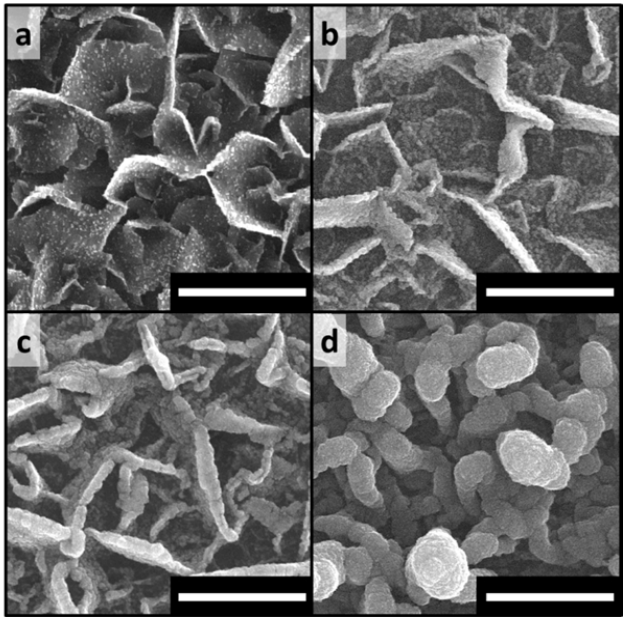

**Figure 1**. SEM images of FLG sheets covered with different amounts of electrochemically deposited palladium (a = 0.64, b = 9.94, c = 22.77, d = 83.32 in µg Pd/cm²). All scale bars are 500 nm.

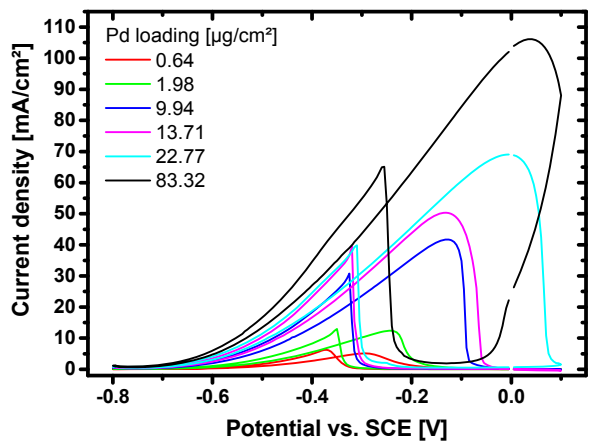

**Figure 2.** Cyclic voltammograms (scan rate: 100 mV/s) in terms of current densities of six Pd-FLG sheet samples of different Pd loadings in 1 M KOH + 1 M ethanol.

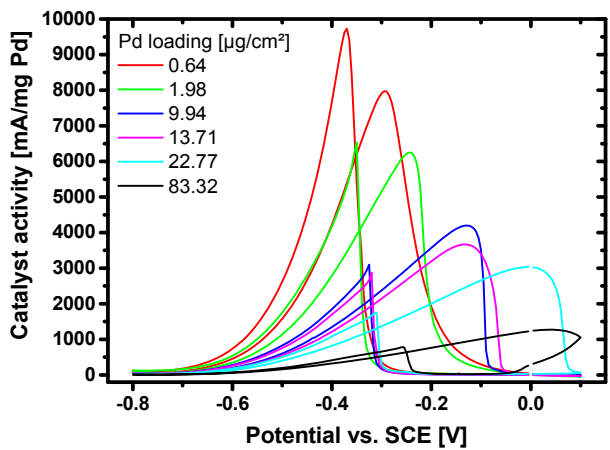

**Figure 3**. Cyclic voltammograms (scan rate: 100 mV/s) in terms of the catalyst activity of the six samples.

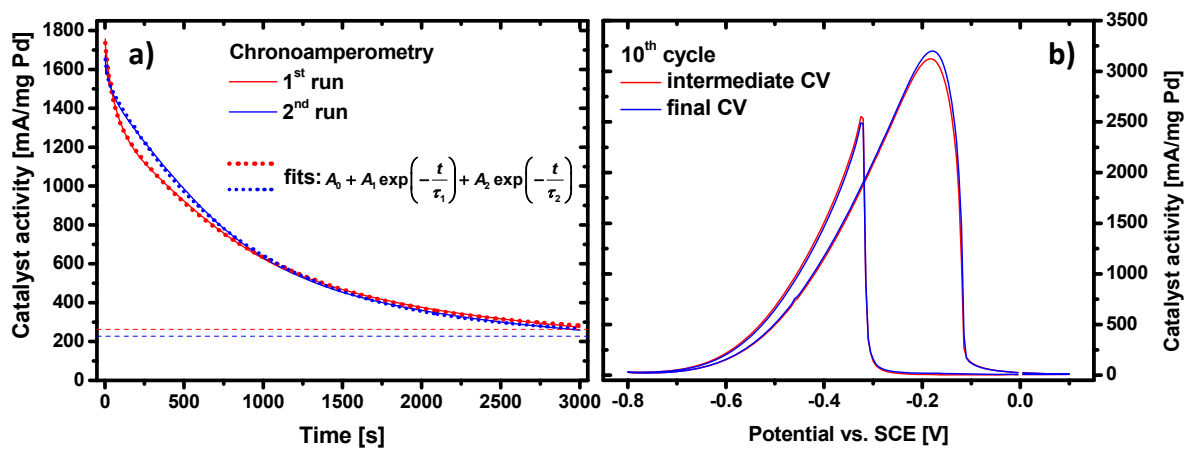

**Figure 4.** (a) The red and blue curves represent two consecutively recorded chronoamperometry traces of the sample with 13.71 µg Pd/cm², each measured for an applied constant voltage of -0.3 V vs. SCE in 1 M KOH + 1M ethanol solution. Dotted lines are biexponential fits. Dashed lines represent the corresponding catalyst activity baselines. (b) The red and blue curve each represents the last of 10 cycles of voltammetry measured directly after the corresponding chronoamperometry traces shown in (a).

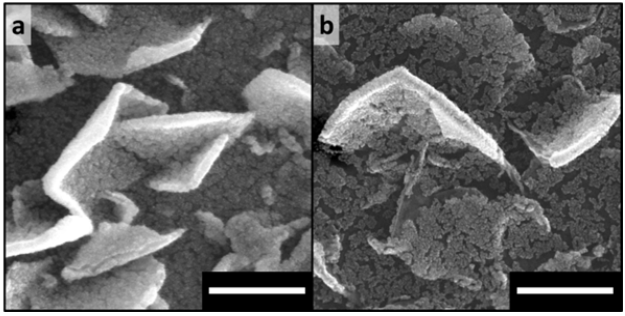

**Figure 5**. SEM images of FLG sheets with a Pd loading of 13.71 µg/cm² (a) before and (b) after all electrochemical measurements. (Scale bars are 300 nm).

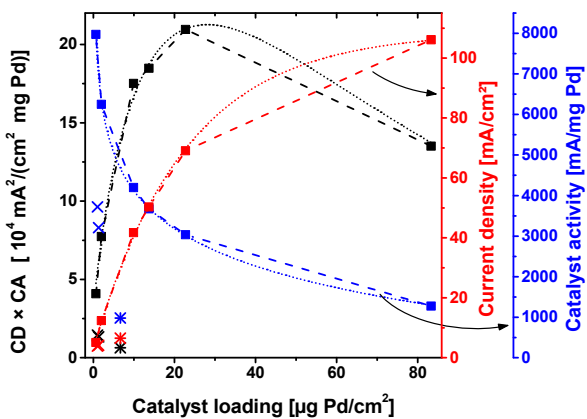

**Figure 6.** Filled squares depict the current density (red), catalyst activity (blue) and their product (CD × CA) vs. the Pd-catalyst loading. Dashed lines connect the data points as guides to the eyes. Red and blue dotted curves are fits to the data points; the product of both curves yields the black dotted curve. The crosses and asterisks represent values of the sputtered-carbon and sputtered-Pd reference samples, respectively.

**Table 1.** Measured current density, catalyst activity, forward anodic peak potential and $I_f / I_b$ - ratio for different Pd catalyst loadings.

| Pd loading (µg/cm²) | current density (mA/cm²) | catalyst activity (mA/(mg Pd)) | forward anodic peak potential (V) vs. SCE | $I_f / I_b$ |
|---|---|---|---|---|
| 0.64 | 5.13 | 7977 | - 0.295 | 1.36 |
| 1.98 | 12.39 | 6249 | - 0.245 | 1.66 |
| 9.94 | 41.72 | 4198 | - 0.130 | 2.61 |
| 13.71 | 50.34 | 3671 | - 0.130 | 2.67 |
| 22.77 | 69.08 | 3034 | - 0.005 | 4.04 |
| 83.32 | 106.11 | 1274 | 0.045 | 3.69 |
| 6.65 * | 6.52 | 981 | - 0.26 | 1.55 |
| 1.05 ** | 3.87 | 3699 | - 0.28 | 1.08 |
| 1.30 *** | 3.78 | 2913 | -0.25 | 1.00 |

*: Results of the reference sample with sputtered Pd (~ 6 nm layer thickness).

**: Results of the reference sample with 5 nm sputtered carbon

***: Results of the reference sample with 10 nm sputtered carbon

**AUTHOR INFORMATION**

**Corresponding Author**

* kipp@chemie.uni-hamburg.de

**ACKNOWLEDGMENT**

T. K. acknowledges funding from the European Union's Horizon 2020 research and innovation programme under the Marie Skłodowska-Curie grant agreement No 656598.

Table of Contents Graphic:

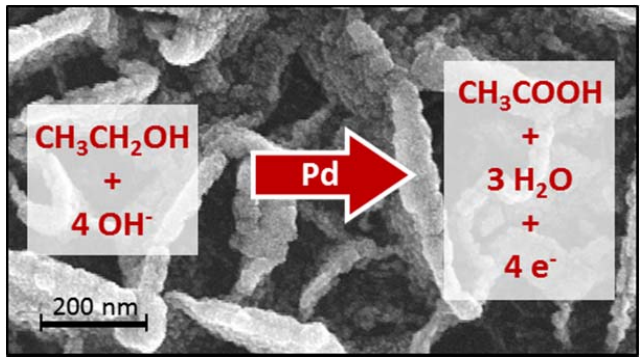